# Audio Bank: A High-Level Acoustic Signal Representation for Audio Event Recognition

Tushar Sandhan, Sukanya Sonowal and Jin Young Choi

Department of Electrical and Computer Engineering,
Seoul National University, Seoul, Republic of Korea 151-742
{tushar, sukanya, jychoi}@snu.ac.kr

**Abstract:** Automatic audio event recognition plays a pivotal role in making human robot interaction more closer and has a wide applicability in industrial automation, control and surveillance systems. Audio event is composed of intricate phonic patterns which are harmonically entangled. Audio recognition is dominated by low and mid-level features, which have demonstrated their recognition capability but they have high computational cost and low semantic meaning. In this paper, we propose a new computationally efficient framework for audio recognition. Audio Bank, a new high-level representation of audio, is comprised of distinctive audio detectors representing each audio class in frequency-temporal space. Dimensionality of the resulting feature vector is reduced using non-negative matrix factorization preserving its discriminability and rich semantic information. The high audio recognition performance using several classifiers (SVM, neural network, Gaussian process classification and $k$-nearest neighbors) shows the effectiveness of the proposed method.

**Keywords:** Audio event classification, non-negative matrix factorization, feature construction, Gaussian process.

## 1. INTRODUCTION

Audio event recognition (AER) or classification is a sub-area of auditory scene analysis that deals with the automatic understanding of audio data without human efforts. This area has received a great attention in recent years, because apart from its straightforward applications in audio retrieval, indexing and automatic tagging, it has the potential to play a pivotal role in perceptually aware interfaces such as computer or robotic assistance in meeting-room, industrial automation and process control, mobile robots working in diverse environments [1], human action recognition [17] and video-surveillance [3], [13]. For example in a video surveillance system, use of an audio information along with the video can enhance the performance of the system for scene understanding especially in dark illumination conditions or in areas outside the camera view, otherwise the underlying activity recognition using only video will become impossible due to absence of perceptual information. In case of a meeting-room assistant robot, detection or classification of certain audio events related to human presence like *chair moving*, *door open*, *keyboard typing* etc. may help the robot to detect and recognize the activities occurring in the room. In these scenarios, the performance of the entire control and automation system highly depends on its ability to correctly classify an audio event.

Most of the previous works in AER [2], [4], [5], [10] make use of standard features such as mel-frequency cepstral coefficients (MFCC), zero-crossing rate, subband energies, short term energy along with their statistical properties like mean, standard deviation and entropy. But these low and mid-level features are limited in the amount of acoustic semantics they can capture, which yield a representation with inadequate discriminative power, hence none of them is a clear winner for robust audio recognition. In this paper, we propose the Audio Bank, a new high-level representation for the audio events, which is semantically rich and highly discriminative in nature. Audio bank explores the set of distinctive audio detectors, which ultimately act like the bases of high dimensional 'audio space' and thereby giving semantically rich high-level representation. Fig.1 gives an overview of the audio bank based feature construction and the AER framework.

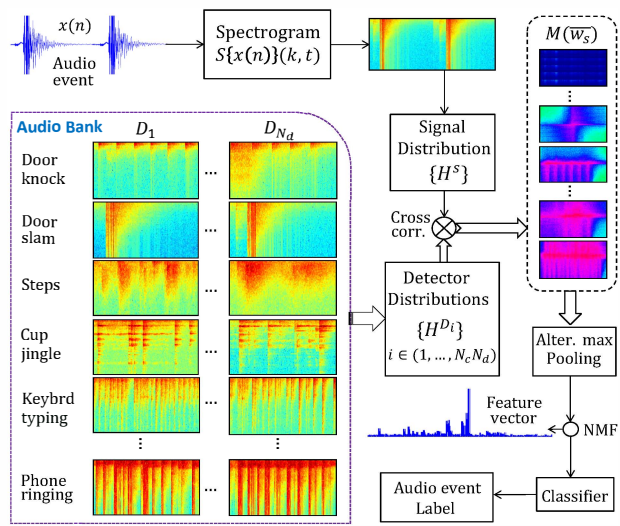

Fig. 1 Overview of the proposed Audio Event Classification (AEC) framework. Audio bank (sec. 2) is the high-level representation of the audio events. It stores a set of audio detectors (sec. 2.2) representing each audio class in the spectrogram space (sec. 2.1). It produces the feature vector (sec. 2.3) by concatenating max-pooled cross-correlation (between detector and signal distribution) patterns from each detector.

82

The low-level and the mid-level features can not comprehensively represent the underlying harmonic structure in the audio data. So conventional audio recognition methods try to uncover the hidden phonic patterns by exploiting the data structure using hidden Markov models (HMM) [22] and Gaussian mixture models (GMM) [20]. These models require a lot of training data. The use of other domain methods for AER, like graph based models [24] and sparse representation [21] is not straight forward and they are also computationally expensive especially for high dimensional feature vectors. Whereas the proposed audio bank feature representation is discriminative enough to produce good recognition results using even a simple classifier like the $k$-nearest neighbors [18].

We have applied the several learning algorithms in order to measure the audio classification accuracy based on these audio bank features. Also several experiments were performed by varying the number of bank detectors and the amount of the training data to assess the dependency of the proposed framework on various parameters. Our major contributions include; proposing a new high-level and robust feature representation using the audio detectors and providing an experimental study of its suitability for the classification task by using several learning algorithms. We use the UPC-TALP database [9] for AER which contains a set of isolated meeting-room acoustic events. It is a balanced dataset [7] i.e. it has enough number of samples per audio class. Experimental results show that the proposed framework provides better AER rate than the conventional feature extraction methods.

## 2. AUDIO BANK REPRESENTATION

Audio signal is composed of intricate phonic patterns which may be repeated several times in the signal. So audio bank tries to exploit this fact and represents an audio as the collected output of many phonic detectors that each produces a correlation pattern. Although, at a glance, audio bank looks related to object bank [16], in detail, the audio classification problem is clearly distinct from the image classification. For the later one, it is evident that image is made up of primary objects like trees, water, mountains, people, buildings etc. So object bank explicitly consists of individual object detectors. But in case of audio signal the atomic phonic patterns from which the audio is formed, are difficult to separate from each other. So we employ a template-based audio detector, where distinctive audio examples themselves serve as the templates. To infuse the robustness to loudness (amplitude), we make use of feature distribution similarity measure (Bhattacharya coefficient) for matching the audio detectors. Invariance to pitch (fundamental frequency of an audio) is achieved via analyzing signal in frequency domain. To account for timbre (differences in pitch quality), we sample distinctive templates from the audio space.

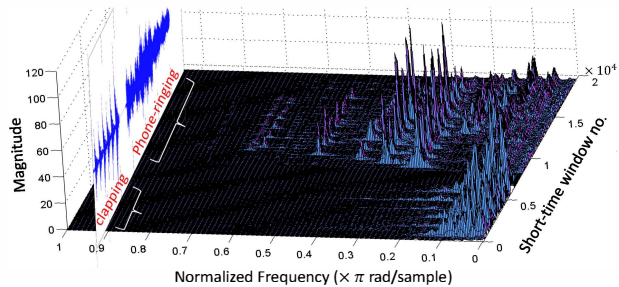

Fig. 2 Two different audio events (*clapping*, *phone-ringing*) and there respective spectrograms. Both signals show different frequency patterns, where *phone-ringing* class shows much higher frequency content.

### 2.1 Feature Extraction

The spectral content of an audio changes over time, so applying the discrete Fourier transform (DFT) over the entire signal does not reveal transitions in spectral content (non stationary signal). But for short periods of time, audio can be considered to be stationary. So our feature space consists of audio spectrogram, which is collection of the power spectrums of short-time signals as follows,

$$S\{x(n)\}(k,t) = \left| \sum_{n=0}^{N-1} x(n+tM)\,w(n)\,e^{-j\frac{2\pi}{N}nk} \right|^2, \quad (1)$$

where $x(n)$ is the audio signal, $w(n)$ is (short-time) analysis window of size $N$ (e.g. Hamming), $k$ is frequency bin index, $t$ is time frame index and $M$ is the framing step (number of samples separating two consecutive frames). Human auditory system is able to distinguish various audio events because they exhibit typical time-varying frequency patterns. Spectrograms and signal waveforms for two different audio events are shown in Fig.2.

### 2.2 Selecting Bank Detectors

Audio bank gives us great deal of flexibility in choosing the kind of bank detectors; indeed different types of detectors (using different features) can be used concurrently. In our implementation, we use distinctive signal spectrogram as the detector due to its evident capability in localizing events from a single template, efficiency (using FFT [23]), pitch analyzing capability and natural interpretation as an audio decomposition into space-time frequencies. In order to capture discriminative audio patterns, the bank detectors must be distinct from each other.

Let $x_i^c$ be the sample from audio (event) class $c \in \{1, \cdots, N_c\}$. Then for each $c$, we choose $N_d$ representatives from $\{S\{x_i^c\}|i=1,2,\cdots\}$ as detectors $\{D_i\}_1^{N_d}$ by using $k$-means clustering [25]. To achieve robustness to timbre, we also sample detectors from *audio space* (audios from the same event but having different sound quality), with constraints that the detector size should be small enough to contain only few occurrences of the underlying audio event. All detectors are stored as spectrograms and thereby producing the bank of size $N_D = N_c \times N_d$.

83

## 2.3 The Audio Bank Feature Vector

Intuitively the audio bank feature vector is nothing but the concatenation of detection responses of all $N_D$ detectors with the signal feature $S\{x\}$. To obtain the detection response in the large audio feature, the detector is placed at each position and the similarities between the frequency distributions (histograms $H^D$ and $H^s$) at the corresponding positions of the detector $D$ and the signal feature $S\{x\}$ are computed. Let this similarity be given by $sim(H^s(\bar{w}_d), H^D(\bar{w}_d))$, where $\bar{w}_d = (k,t)$ ranges over frequency-time support of the detector. The global similarity match measure, $M(\bar{w}_s)$, at each position $\bar{w}_s = (k,t)$ of the signal feature is obtained by summing the individual similarity measure across the detector as,

$$M(\bar{w}_s) = \sum_{\bar{w}_d} sim(H^s(\bar{w}_d), H^D(\bar{w}_d - \bar{w}_s)). \quad (2)$$

We use the Bhattacharyya coefficient [6] as the similarity measure, because of two reasons; first, it bounds the values between 0 and 1, with 1 indicating complete match; second, it yields to efficient implementation. The Bhattacharyya coefficient for two histograms $H^1$ and $H^2$, each with $B$ bins and $b$ as the bin index, is defined as,

$$sim(H^1, H^2) = \sum_{b=1}^{B} \sqrt{H_b^1, H_b^2}. \quad (3)$$

Now after inserting measure (3) into global match measure (2) and rearranging the summation orders produces,

$$M(\bar{w}_s) = \sum_b \sum_{\bar{w}_d} \sqrt{H_b^s(\bar{w}_d)} \sqrt{H_b^D(\bar{w}_d - \bar{w}_s)}, \quad (4)$$

which is the cross-correlation between the individual bins of the histograms as, $M(\bar{w}_s) = \sum_b \sqrt{H_b^s} \star \sqrt{H_b^D}$. These 2-dimensional correlations are computed efficiently in the frequency domain using the convolution theorem of the Fourier transform [8], where the computationally expensive correlation operations are exchanged for relatively simpler pointwise multiplications as follows,

$$M(\bar{w}_s) = F^{-1}\left\{\sum_b F\{\sqrt{H_b^s}\} F\{\sqrt{\widetilde{H}_b^D}\}\right\}, \quad (5)$$

where $\widetilde{H}_b^D$ is the flipped detector distribution with $F\{\cdot\}$ and $F^{-1}\{\cdot\}$ denoting the Fourier transform and its inverse respectively. The Fourier transforms are realized efficiently by using FFT algorithm. The audio bank feature vector is constructed from the global correlation pattern, $M(\bar{w}_s)$, by using alternate max-pooling operation. In this operation, the output correlated pattern is subsequently divided into $\{2^0, 2^2, 2^4\}$ equal parts and then maximum values from all the parts are concatenated one after another to yield $N_d \times N_c \times 21$ dimensional feature vector for each input audio signal. Alternate max-pooling operation (shown in Fig.3) efficiently and compactly captures the underlying variations in the correlation output. Thereby it tries to find the maximally correlated frequency patterns.

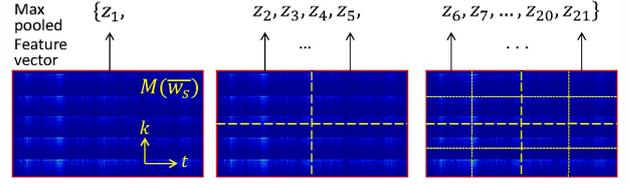

Fig. 3 Alternate max-pooling operation extracts feature vector from the correlation output of each detector.

## 2.4 Non-Negative Matrix Factorization (NMF)

NMF is a signal representation method for noise-robust feature extraction in the reduced dimension [11]. It aims to minimize the distance between the original signal and its approximation. Given a non negative $m \times n$ matrix $X$, NMF decomposes it into the non-negative $m \times k$ matrix $W$ and the non-negative $k \times n$ matrix $H$ that minimize the following reconstruction error,

$$J(W, H) = \|X - WH\|_F^2, \quad (6)$$

where $\|\cdot\|_F$ denotes the Frobenius norm of a matrix. The column vectors of $W$ are the basis vectors and the values along the column in $H$ denote the contribution of the column vectors of $W$ or in other words, it represents the decomposition of the signal values on the basis matrix $W$. NMF thus finds the 'parts' based, linear representations of non-negative data as only additive combinations of the bases are allowed. $W$ can be learned according to (6) using the algorithm proposed in [15] and then $H$ is learned with a fixed $W$. In our work $X$ is the matrix of max-pooled feature vectors (see Fig.3). The decomposition of features on the basis matrix is then treated as the new coded features, which are the input to the classifiers.

## 3. EXPERIMENTS

### 3.1 Dataset Description and Experimental Setting

The UPC-TALP dataset [9] contains 14 classes of audio events that occur in a meeting room environment. The events are recorded such that they do not overlap in the time domain and the recording is done at 44.1kHz sampling frequency. Table 1 shows the 14 audio classes present in the database along with their annotating labels. In our work, we did not use the class "unknown" for training and evaluation, because it is not of much significance and contains occurrences of silence and noise. The classes "door open" and "door close" perceptually sound similar (because both produce the same door slam sound), so we mixed them to make the unified class "door movement" having total 121 samples. So we were left with 12 classes over which we report our AER results.

For the feature extraction process, the audio samples were down-sampled by $\frac{1}{4}$ and 256-point discrete Fourier transform (DFT) was taken with a frame overlap of 50%

84

Table 1  Audio events in the UPC-TALP database with their number of audio samples and total duration (sec).

| Event Name | # Samples \ Time | Event Name | # Samples \ Time |
|---|---|---|---|
| Door knock (kn) | 50 \ 64 | Key jingle (kj) | 65 \ 146 |
| Door open (do) | 60 \ 66 | Keyboard typ. (kt) | 66 \ 194 |
| Door close (ds) | 61 \ 78 | Phone ring (pr) | 116 \ 315 |
| Steps (st) | 73 \ 250 | Applause (ap) | 60 \ 212 |
| Chair move (cm) | 76 \ 216 | Cough (co) | 65 \ 85 |
| Cup jingle (cj) | 64 \ 187 | Laugh (la) | 64 \ 118 |
| Paper work (pw) | 84 \ 300 | Unknown (un) | 126 \ 89 |

to compute the spectrograms. For the AER analysis using various classifiers, 60% samples from each class were randomly assigned as the training data and the rest 40% samples were used for testing. No audio from the audio bank is used again for testing stage. Extensive experiments were performed (described in following sections) to validate the robustness of the proposed features.

### 3.2 Audio Bank with Different Classifiers

We applied several classifiers individually on the audio bank features to evaluate their suitability and discriminability for AER. We have used, $k$-nearest neighbor (kNN), one-vs-all SVM (SVM-A), one-vs-one SVM [19] (SVM-O), Gaussian process [12] (GP) and Neural Network [26] (NN) for classification. For the kNN, $k$ was set to 5. For SVM-A we used RBF kernel with $C = 150, \sigma = 75$, while $C = 100, \sigma = 60$ for SVM-O. These parameters were determined by 5 fold validation.

GP is completely specified by its mean and covariance function. The mean function for the GP classifier was set as a constant and an isotropic squared exponential kernel was used as the covariance function. We implemented the GP from the GPML toolbox [14], which learns the GP hyper-parameters and predicts the class labels. As GP classifier is essentially a binary classifier, we used one-vs-one approach to extend its functionality for multi-class classification. For Neural Network, the number of hidden layers was arbitrarily set to 300.

For each classifier, we performed the recognition experiment 5 times i.e. for each of the 5 runs the data were randomly assigned into the training and test set and the classifier was subsequently applied. Fig.4 shows the average AER accuracy obtained using audio bank features, the whisker at the top of each bar denotes the standard deviation. We can see that the AER performance is fairly good for a simple classifier like kNN, whereas for more complex classifiers like SVM and NN, the performance is significantly better. In case of GP, however the accuracy suffers possibly due to the high feature dimensionality.

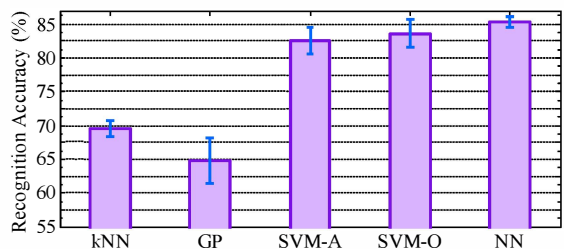

Fig. 4  AER by Audio Bank using different classifiers.

Table 2  Simulation time of different classifiers for the audio bank features averaged over the 5 experimental runs, where the size of the audio bank is kept fixed to $N_D = 48$

| Classifiers | k-NN | SVM-A | SVM-O | GP | NN |
|---|---|---|---|---|---|
| Simulation Time (sec) | 0.49 | 0.89 | 46.44 | 180 | 10.72 |

**Computational time**: Table 2 shows the total simulation time taken for testing and training in each of the above mentioned classifiers. As evident, kNN requires the least time due to its simplicity. SVM-O and GP on the other hand are comparatively much slower as these classifiers are implemented in one-vs-one fashion where the computational complexity scales quadratically with the number of classes. The simulation time for SVM-A is quite less (comparable to kNN), while NN classifier is moderately slower. Thus, if we compare all the 5 classifiers; according to the simulation time and the recognition accuracy, then SVM-A is a clear winner as it performs considerably well without much compromise on speed.

**Exploring kNN classifier**: The dependency of AER performance on the value of parameter $k$ in case of the kNN classifier is assessed by varying $k$ from 1 to 100 in intervals of 10. For each $k$, the recognition performance was recorded 10 times and the average performance is reported in Fig.5 along with the corresponding standard deviation indicated by whisker. From the figure we see that the recognition performance drops with increase in $k$. Thus, lower the $k$ values, the better the AER accuracy.

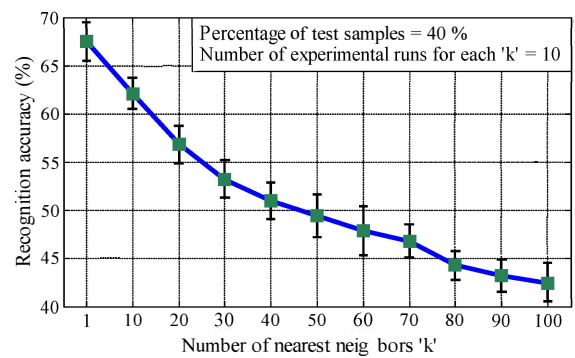

Fig. 5  Average recognition accuracy of kNN classifier using audio bank features, with nearest neighbors ($k$) varying from 1 to 100. Mean and standard deviation is denoted by marker and the whisker respectively.

**Exploring Neural Network**: We also analyzed the effect of the number of hidden layers in the NN classifier on the recognition performance. The number of hidden layers was varied from 10 to 1000 and the recognition performance for each hidden layer value was measured by averaging over 10 runs, similar to the kNN experiment.

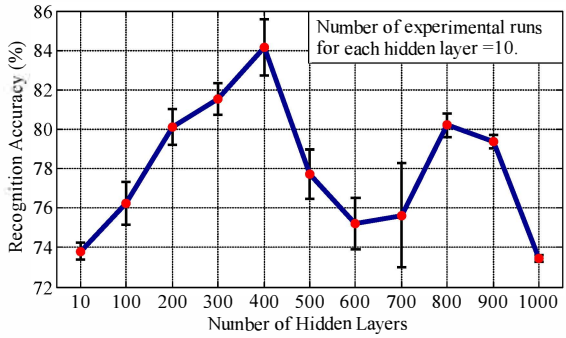

Fig. 6 Average recognition accuracy of Neural Network classifier using audio bank feature vectors, with number of hidden layers varying from 10 to 1000. The circular marker denotes the average obtained over 10 runs and the whisker denotes the standard deviation.

Fig.6 shows the AER accuracy obtained for each hidden layer value. It shows that if the number of hidden layers is in between 300 and 400, the recognition performance is considerably good. Also for other number of hidden layers, the performance does not degrade significantly.

**3.3 Comparison with Other Methods**

Audio bank (size $N_D = 48$) features are compared with 3 sets of conventional features using the same SVM-O classifier. Feature 'set-A' consists of 32 dimensional (32-D) log-filter bank energies, the zero crossing rate (ZCR) and the signal concatenated into a 34-D feature vector. Feature 'set-B' consists of 2050-D feature vector constructed by using FFT coefficients [8], ZCR and signal energy. And feature 'set-C' consists of 13-D MFCC coefficients [4]. Table3 shows the average AER rate achieved over 5 runs of the experiment. The performance of the audio bank features is significantly better than the conventional features. Its per class performance is portrayed in the form of confusion matrix in Fig.7. High diagonal entries signify the interclass discriminability of the proposed features. The method slightly confuses for 'chair moving' class, where it miss-classifies some of the entries to 'cup jingle'. The method achieves perfect recognition rate for 'steps' and more than 90% accuracy for 'door knock', 'paper work' and 'phone ring' classes.

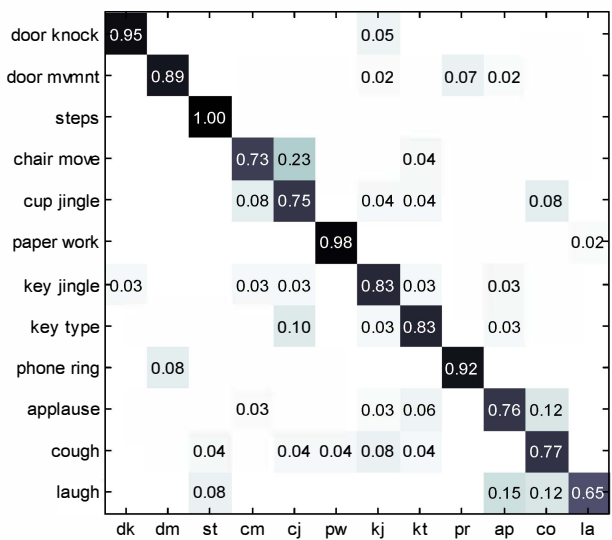

Fig. 7 Confusion matrix (AER rate) using Audio Bank.

Table 3 Conventional feature sets vs Audio Bank feature.

| Methods | set-A [10] | set-B [8] | set-C [4] | Audio Bank |
|---|---|---|---|---|
| **AER rate** (%) | 51.44 | 44.16 | 73.38 | 84.59 |

**3.4 Training Data Size Variation**

Amount of training data plays a vital role in deciding the classification performance. More the training data, better will be the generalization performance. So the amount of training data from each class was varied from 10 to 90% (correspondingly test data per class became 90 to 10%); then the resulting AER performance is summarized in Fig.8. It also shows the instability of NN classifier (due to high dependency on parameter initializations). The performance of the SVM-A and SVM-O gradually increases as the training data increases. We can see that, even 40% of training data per class gives more than 80% AER rate. This shows ability of the proposed features for good discrimination and class generalization.

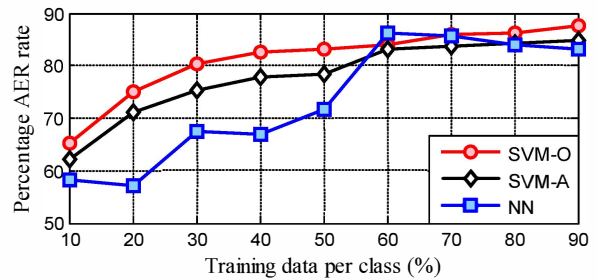

Fig. 8 Audio event recognition (AER) rate for three classifiers with varying the amount of training data.

**3.5 Audio Bank Size Variation**

The proposed framework is very general and can easily adapt to different AER settings by simply adding more new distinctive detectors to the existing bank. However it is not obvious that a larger bank necessarily gives better performance, because as the number of bank detectors ($N_D$) increases, dimensionality of the audio bank feature vector also increases and AER rate may hinder due to the curse of dimensionality. To assess these effects we performed AER by gradually varying the bank size. Audio bank consists of distinctive $N_d$ detectors from each audio class (see section 2.2). We varied the bank size from 3 to 120. The dependency of AER rate (using 60% of training data from each class) with the bank size is shown in Fig.9. For NN classifier, the slight decay in AER rate with big bank is observed due to the curse of dimensionality and instability of the NN classifier (see section 3.3). Tiny bank ($N_D$ = 3 or 6) had smaller amount of detectors than total classes (12). So some classes were remained unexpressed by the resulting feature vector. Thus tiny bank performed poorly. But even the small bank ($N_D$ = 12, one detector for each class) has produced comparable accuracy (80%) to big bank ($N_D > 50$). This surprising stability of the audio bank on its size can be attributed to its ability to represent event in the audio space and thereby producing a high-level representation which is also group sparse in nature. Because for responding to a particular class of audio event, only the detectors that are closely related to it, will produce high detection response; whereas the semantically remote detectors will remain in the dormant state. Thus even after using the small bank, the resulting group sparse and semantically rich feature is powerful enough to produce high AER performance.

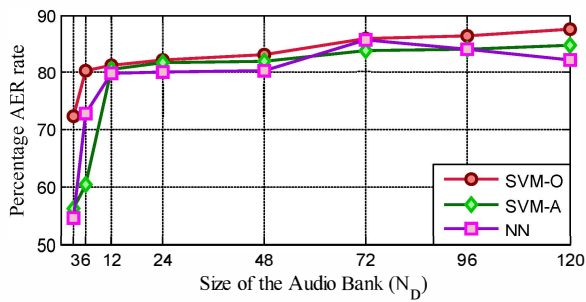

Fig. 9 Analyzing the effect of bank size on AER rate. With notations defined as, tiny bank (detectors $N_D = 3, 6$), small bank ($N_D = 12$), big bank ($N_D > 50$).

## 4. CONCLUSION

We proposed Audio Bank, a new high-level efficient audio representation for AER, aimed at exploring the underlying harmonic structure present in the audio data. It is comprised of distinctive audio detectors representing each audio class in frequency-temporal space. It produces superior features as compared to low-level features in discriminating audio events, by appropriately pooling a large set of smaller audio detectors. Feature stability on the bank size and high AER performance using several classifiers (SVM, neural network, Gaussian process and $k$-NN) shows the effectiveness of the proposed method.